\documentclass[epjCONF]{svjour}
\usepackage{graphicx}
\usepackage{graphics}
\usepackage[varg]{txfonts} 
\usepackage[latin1]{inputenc}

\newcommand{\apj}{Ap. J.}

\newcommand{\mnras}{Mon. not. RAS.}

\newcommand{\nat}{Nature}

\def\lesssim{\mathrel{\hbox{\rlap{\hbox{\lower4pt\hbox{$\sim$}}}\hbox{$<$}}}}
\def\gtrsim{\mathrel{\hbox{\rlap{\hbox{\lower4pt\hbox{$\sim$}}}\hbox{$>$}}}}
\session-title{Tidal Distuption Workshop}
\begin{document}
\title{Time Scales in Tidal Disruption Events}
\author{Tsvi Piran\inst{1}\fnmsep\thanks{\email{tsvi@phys.huji.ac.il}} \and Julian Krolik\inst{2}\fnmsep\thanks{\email{jhk@jhu.edu}}  }
\institute{Racah Institute for Physics, The Hebrew University, Jerusalem,  91904, Israel \and Physics and Astronomy Department, Johns Hopkins University, Baltimore, MD 21218, USA}
\abstract{
We explore the temporal structure  of tidal disruption events  pointing out the corresponding  transitions in the lightcurves of the thermal accretion disk and of the jet emerging from such events.  
The hydrodynamic time scale of the disrupted star is the minimal time scale
of building up the accretion disk and the jet and  it sets a limit on the rise time. 
This suggest that {\it Swift} J1644+57, that shows several flares with a rise time as short as  a few hundred seconds could not have arisen from a tidal disruption of a main sequence star whose hydrodynamic time is a few hours. The disrupted object must have been a white dwarf. 
A second important time scale is the Eddington time in which the accretion rate changes form super to sub Eddington. It is possible that such a transition was observed in the light curve of  {\it Swift} J2058+05. If correct this provides intersting constraints on the parameters of the system.    } 
\maketitle
\section{Introduction}
\label{intro}
 
Tidal disruption events (TDE) of s stellar mass object by a massive black hole are  a classical transient phenomenon that involves numerous time scales: (i)  the gravitational time scale of the massive black hole (ii) the orbital period at the tidal radius, which equals the hydrodynamic time scale of the disrupted star, (iii) the orbital period at the innermost semi-major axis of the disrupted stellar material on which the accretion rate peaks, (iv) the transition from super so sub Eddington accretion and (v) the transition from radiation dominated inner disk region to gas pressure dominated.  
For typical parameters these values range from a few dozen seconds to several years, all but the second depend strongly on the black hole's mass. If interpreted correctly, they provide invaluable information concerning the  system. We examine these times scales and their implications to the interpretation of the observations of {\it Swift} J1644+57 and J2058+05.

\section{Dynamical Time Scales}
\label{sec:1}

The shortest time scale expected in a TDE is the orbital period, $P_{\rm orb}$, of the tidal radius, $R_T$. It  is comparable to the hydrodynamic time scale in which matter can be extracted from the disrupted star. Note that $t_{hyd}$ is independent of the disrupting black hole's mass:
\begin{equation}
t_{hyd} \simeq P_{\rm orb}(R_T) \simeq 10^4~ {\rm sec} ~ (k/f)^{1/4} {\cal M}_*^{(1 -3\xi/2)} ,
\end{equation}
where ${\cal M}_*$ is the mass of the star in solar units. 
$k/f \simeq 0.02$ for radiative stars, but is $\simeq 0.3$ for convective ones (see \cite{phinney89,kp11,kp12} for details).   We have approximated the main
sequence mass-radius relation by $R_* \approx R_\odot M_*^{(1-\xi)}$;
$\xi \simeq 0.2- 0.4 $ \cite{kipp94,kp12}. 

The  disrupted star is spread out and the 
most bound matter, at $a_{min}$,  may have a binding energy as great as $\sim GM_{BH}R_*/R_p^2$, where $R_p\equiv \beta R_T$ is the pericenter of the initial orbit and $\beta$ is the ``penetration factor". The corresponding
period is:
\begin{equation}\label{eqn:porbamin}
t_0  \simeq P_{orb}(a_{min}) \simeq1.5 \times 10^6 ~{\rm sec} ~{\cal M}_*^{(1-3\xi)/2} M_{BH,7}^{1/2}(k/f)^{1/2} \beta^{-3}. 
\end{equation}
The accretion rate peaks on this time scale \cite{rees88,phinney89}.
Assuming  that the disrupted stellar derbies have a uniform distribution in orbital binding
energy per unit mass, matter returns to $\sim R_p$ and accrets onto the black hole
at a rate  $dM/dt \propto (t/t_0)^{-5/3}$. Note that $t_0 \simeq (M_{BH}/{\cal M}_*)^{1/2} P_{\rm orb}(R_T)\gg P_{\rm orb}(R_T)$.
Intersecting  streams
could lead to a conversion of orbital energy to heat, diminishing these orbital
periods, but in no case we would expect time scales shorter than $t_{hyd} \simeq P_{\rm orb}(R_T)$. 
Numerical simulations of the disruption of a main sequence star by a $10^6 M_\odot$ black hole \cite{Ayal+00} show a rather continuous accretion rate with a rise time of a few times $10^5$ sec and an overall duration of a few times $10^6$ sec, as expected from these analytic estimates.

\section{Implication for {\it Swift} J1644+57}
\begin{figure}
\vskip -.7cm
\includegraphics[width=0.4\textwidth,angle=-90]{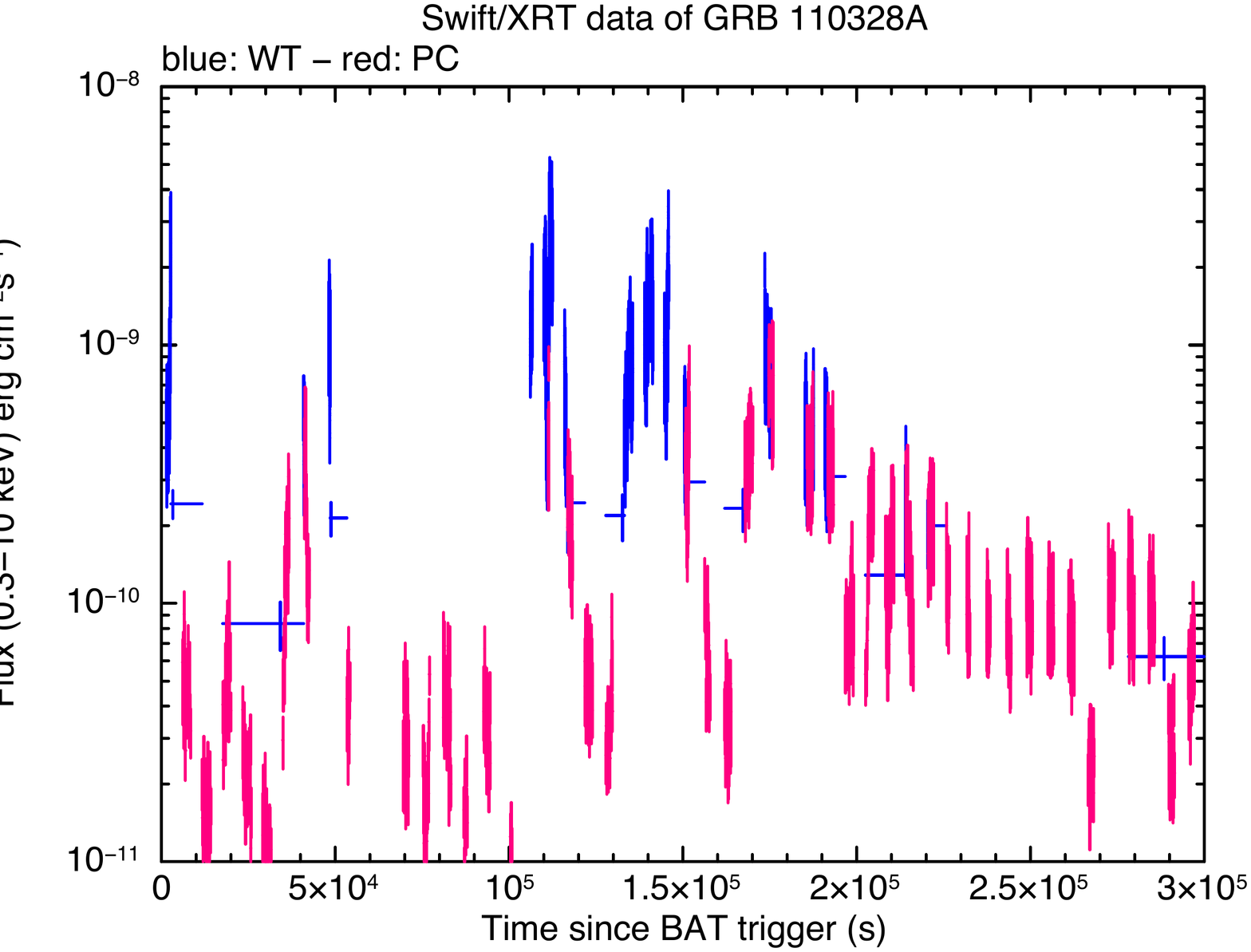}
\includegraphics[width=0.4\textwidth,angle=-90]{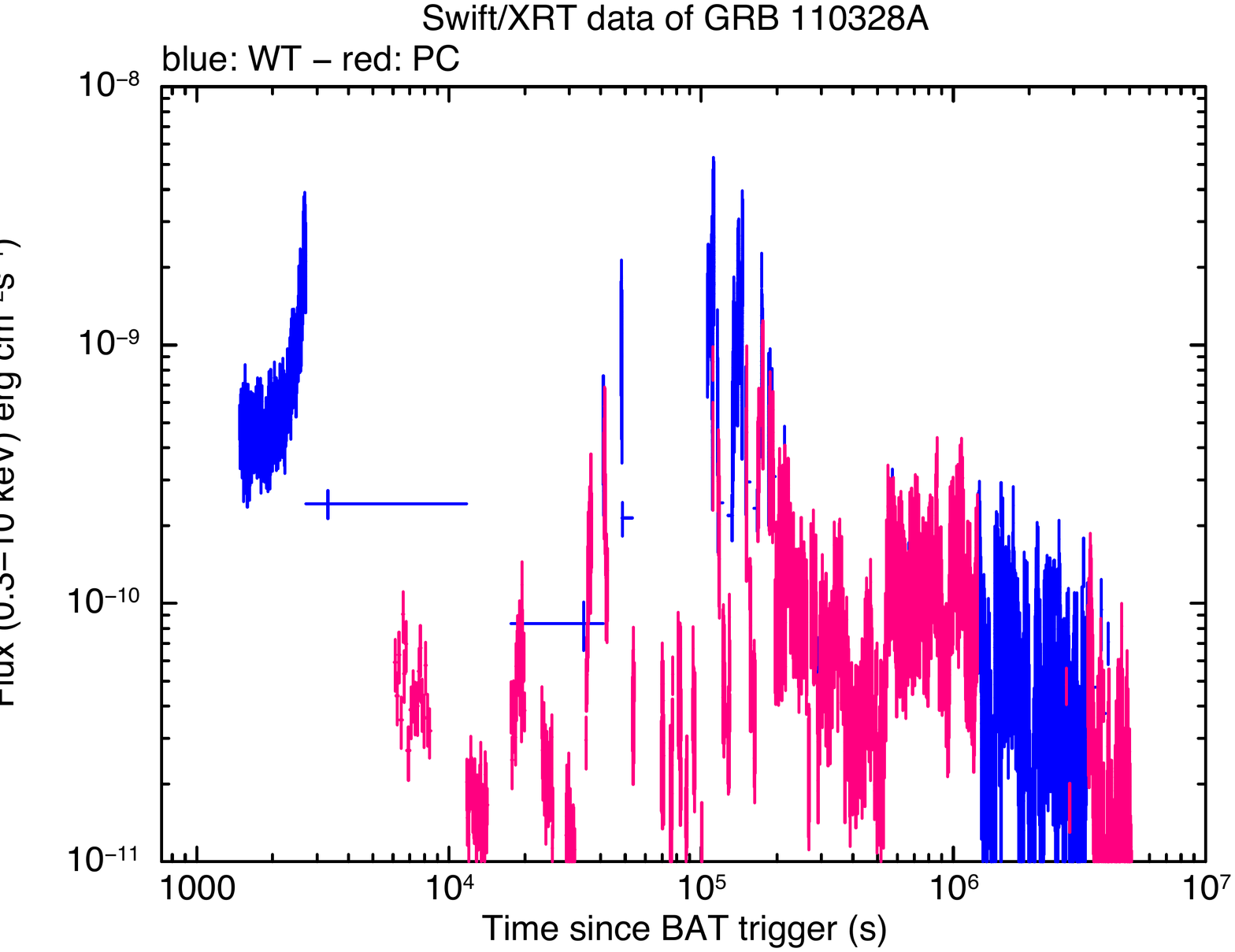}
\vskip -.4cm
\caption{Long-term Swift XRT light curves in the 3--10 keV band (blue: WT, red: PC).
(left) Linear in time representation of the first 300,000~sec, illustrating the
recurring brief flares that gradually widen.  (right)
Logarithmic in time representation of the entire light curve as of 29 August 2011,
five months after activity began. From \cite{kp11}.}
\vskip-.6cm
\label{fig:longterm}
\end{figure}

The flaring x-ray source {\it Swift} J1644+57 \cite{burrows+11}  resembled   initially
a classical $\gamma$-ray burst. However, its  repeated  extremely short  flares 
separated by a few thousand seconds and its long lasting emission revealed 
that this is not the case. The location of the burst
at the center of its host galaxy led  to the suggestion  
that it is  a TDE driven jet \cite{bloom+11,levan+11}. 
Overall,   {\it Swift} J1644+57 reveals a number of characteristic times: rise-times as short
as $\sim 100$~sec; flare durations $\sim 1000$--10000~sec; quiescent periods $\sim 5 \times 10^4$~sec
long; a transition to a smooths flow around $10^5$ sec; and a total event duration of more than $10^7$~sec (see Fig. \ref{fig:longterm}).

It is hard to reconcile these observed timescales with the dynamical estimates.  The rise
time in the consensus model, a  main sequence star disrupted by a $\sim 2 \times 10^7 m_\odot$ black hole, is said to reflect the light-crossing time across the
black hole's horizon,  $t_{g} = {G M_{BH} }/{c^3} = 50 ~{\rm sec}~ M_{BH,7}$.  But it is not clear what dynamics link that quantity to
triggering a flare, nor is there any natural explanation for the flare duration. Moreover, as it takes 
$t_{hyd} \sim P_{\rm orb}(R_T) \simeq 10^4~ {\rm sec}$ to drain the mass from the disrupted star, it is not clear how a rise time significantly shorter than that arises. 
Furthermore,   the timescale at which the flares merged
into a smoother lightcurve and the power-law  decay begins, $\sim 10^5$ sec, appears to be at
least one order of magnitude shorter than expected $t_0$.
These difficulties have led us \cite{kp11} to suggest that  {\it Swift} J1644+57 arises due to the disruption of a white dwarf (WD) by a $\sim 10^4$--$10^5 M_{\odot}$ black hole. Furthermore, the WD  it is not disrupted all at once, but instead it loses pieces of itself in several passes before dissolving. 
This suggestion is
motivated by the fact that the fundamental timescale of a tidal disruption is dictated by
the mean density of the star; the greater density of the WD corresponds to a hydrodynamic time scale of a few seconds which makes it much easier to achieve the short timescales of this event. While this resolves many of the problems associated with the light curve this solution requires a relatively light disrupting black hole,  below typical masses observed in galactic centers and below the  $\simeq 2 \times 10^7M_{\odot}$ estimate based on the   $M_{BH}$--bulge luminosity correlation, \cite{burrows+11}.

\section{Accretion Time Scales} 
Once the matter returns to the vicinity of $R_p$ it is captured
into an accretion disk whose inflow time $t_{\rm in} \ll P_{\rm orb}(a_{\rm min})$.
As the matter moves inward through this disk, there is local dissipation of the
conventional accretion disk variety, and the heat is radiated in the usual quasi-thermal
fashion.  At the peak accretion rate estimated by \cite{lkp09} the luminosity would
be:
\begin{equation}\label{eq:charlum}
L_{\rm peak}\sim 3 \times 10^{46} \hbox{~erg/s} ~(\eta/0.1) {\cal M}_*^{(1+3\xi)/2}
    M_{BH,7}^{-1/2}(k/f)^{-1/2}\beta^3 \simeq 25 \beta^3(\eta/0.1){\cal M}_*^{(1+3\xi)/2}M_{BH,7}^{-3/2} ~ L_{Edd} , 
\end{equation}
for radiative efficiency $\eta$ (see also \cite{ulmer99,strubbe09})\footnote{The Eddintgon luminosity ratio is for a radiative star, for a convective star it is lower by a factor of 4.}.
In a very narrow mass range\footnote{We have normalized most parameters to their canonical values.}: $ 9 \times 10^7 m_\odot ~ {\cal M}_*^{(1+3\xi)/3} \lesssim  M_{BH} \lesssim 1.3 \times 10^8 m_\odot~ {\cal M}_*^{(2 - 3 \xi)/2} $ the luminosity won't exceed Eddington. If the black hole is more massive it  swallows the  star before  tidally disrupting it. 

\begin{figure}
\vskip -.5 cm
\includegraphics[width=0.5\textwidth]{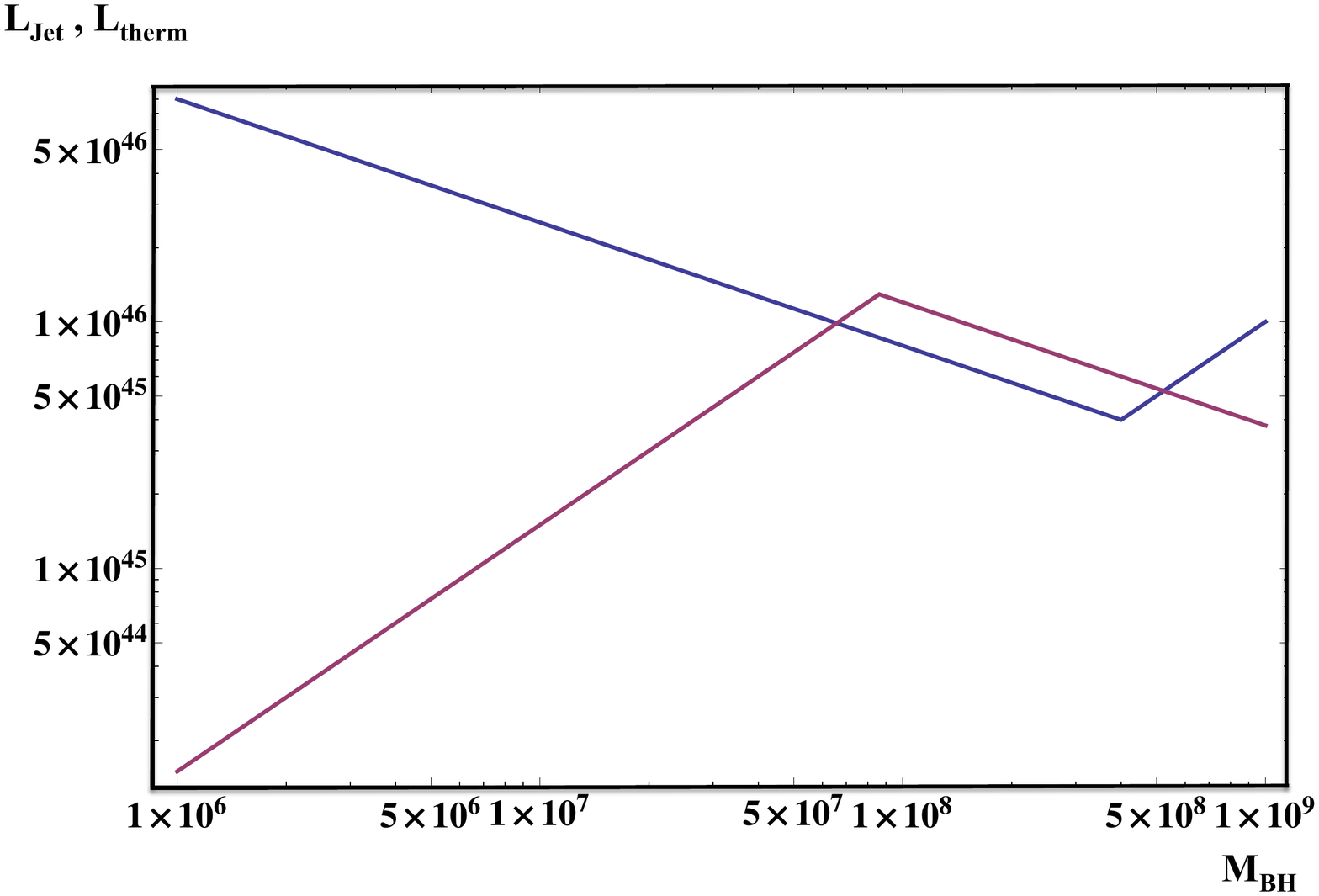}
\includegraphics[width=0.5\textwidth]{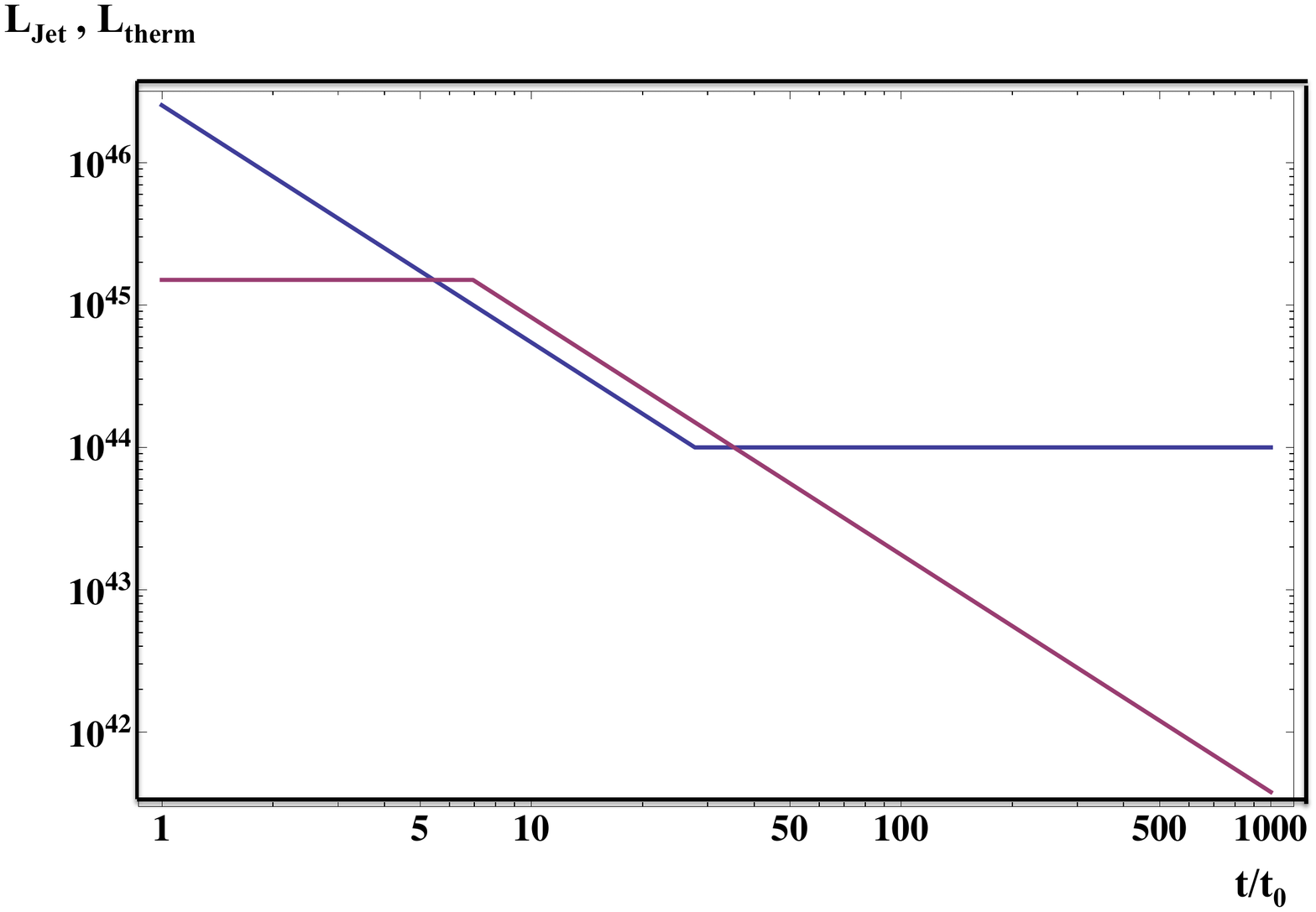}
\vskip -.6cm
\caption{(a) Left: Peak jet power (blue) and thermal (red) luminosity as a function of black hole
mass.  (b) Right: Jet power (blue) and thermal (red) luminosity as functions of time for $M_{BH}
= 1 \times 10^7 M_{\odot}$.  The Eddington timescale $t_{\rm Edd}$ for these parameters is $7 t_0$.   From \cite{kp12}.}
\vskip-.5cm
\label{fig:L-MBH}
\end{figure}

The accretion can lead also  to the formation of a jet powered by a Blandford-Znajek process \cite{giannios11}. One 
 can estimate the power of this jet as \cite{kp12}: $L_{jet} \approx c p_{\rm mid} r_g^2$,
where $p_{\rm mid}$ is the mid pressure at the inner region of the disk and $r_g$ is the gravitational radius of the black hole  (see Fig. \ref{fig:L-MBH}).

As noted already by Ulmers \cite{ulmer99} the accretion is initially super Eddington. 
The transition to sub-Eddington takes place at: 
\begin{equation}
t_{Edd} \simeq 7 \beta^{-6/5} (\eta/0.1)^{3/5} M_*^{3(1+3\xi)/10} M_{BH,7}^{-9/10} t_0 \simeq 10^7 ~{\rm sec} ~  \beta^{-6/5} (\frac{\eta}{0.1})^{3/5}  ( \frac{k/f}{0.02} )^{1/2} M_s^{4-3 \xi)/5} M_{BH,7}^{-2/5} .
\end{equation}
From the disk we expect for $t_0<t< t_{Edd}$  a roughly constant bolometric luminosity $L_{disk} \approx L_{Edd}$ and a decreasing luminosity, $\propto t^{-5/3}$ at later times \cite{ulmer99,strubbe09,LR11}. 
We \cite{kp12} have found an accompanying transition at the jet luminosity at $t_{Edd}$. 
For $t < t_{Edd}$ the jet power is proportional to the accretion rate and it
decrease like $t^{-5/3}$. On the other hand for $t> t_{Edd}$, the disk becomes thin and radiation dominated. In such a case the pressure in the innermost region of the disk is independent of the accretion rate \cite{moderski96}. This leads to a roughly constant jet luminosity in this regime. This constant phase will continue until the inner regions of the disk becomes gas dominated leading to a rapid decrease in the jet luminosity.

Figure~\ref{fig:L-MBH}b gives a schematic view, beginning at the time of peak accretion
rate, of what might be expected in terms of the light curves for the jet power (before
allowance for beaming and radiative efficiency) and the thermal disk luminosity.  For the
parameter values chosen ($M_{BH,7} = 1$, all other scaling parameters unity),
$L_{\rm jet}$ falls to the level of $L_{\rm therm}$ at almost the same time, $t \simeq 7 t_0$,
as $L_{\rm therm}$ enters the sub-Eddington regime and also begins to decline.  From that
time to $t \simeq 30 t_0$, both fall together, maintaining similar power levels.  Finally,
after $t \simeq 30t_0$ (i.e., a time larger by $\eta^{-3/5} \simeq 4$ than the time at which the
thermal luminosity begins to decline), the jet luminosity stabilizes, while
$L_{\rm therm}$ continues to fall.

\section{Implication for {\it Swift} J2058+05}
\begin{figure}
\vskip -1.5cm 
\begin{center}
\includegraphics[width=0.5\textwidth,angle=90]{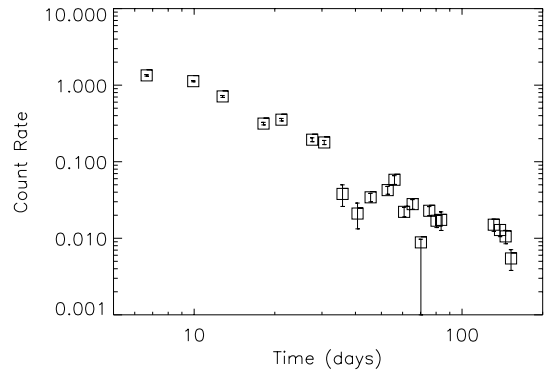}
\end{center}
\vskip -1.4cm
\caption{Long-term Swift XRT light curves in the 0.5--10 keV band for J2058+05
as of 2 November 2011, combining WT and PC data
(data drawn from http://www.swift.psu.edu/monitoring).
Many error bars are smaller than the associated plot symbols.  Where the error bar extends
indefinitely downward in this logarithmic plot, it represents an upper bound. From \cite{kp12}.
\label{fig:lc}}
\vskip-.5cm
\end{figure}

{\it Swift} J2058+05  \cite{cenko+11} might be an example demonstrating these effects. 
The x-ray lightcurve is shown in Figure~\ref{fig:lc}.
For $\simeq 10$~d, its flux stayed nearly constant; for the next
three months, it declined $\propto t^{-2}$.  Starting at $t \simeq 70$--100~d, the
decline appears to have become much more shallow.  In \cite{kp12} we compare these features in the 
light curve with the predictions of the model. If we identify the first break
in the lightcurve with $P_{\rm orb}(a_{\rm min})$ and the second one as $t_{Edd}$ then  under reasonable assumptions we can  estimate the mass of the black hole,  $M_{BH} \sim 7 \times 10^7 m_\odot$, as well as other physical parameters of the system (see \cite{kp12} for details).  

\section{Conclusions}
We have explored the various time scales in TDEs and have shown their implications for two TDE candidates, demonstrating the power of a detailed temporal analysis of such events. 

{This work was partially supported by NSF grants AST-0507455 and
AST-0908336 (JHK) and by an ERC advanced research grant  (TP).  }

\end{document}